# Determination of the magnetic phase diagram in the H-T plane for a sigma-phase $Fe_{47}Mo_{53}$ alloy

Stanisław M. Dubiel[1*], Maria Bałanda[2] and Israel Felner[3]

[1]AGH University of Science and Technology, Faculty of Physics and Applied Computer Science, PL-30-059 Kraków, Poland, [2]Institute of Nuclear Physics, Polish Academy of Sciences, PL-31-342 Kraków, Poland, [3] Racah Institute of Physics, The Hebrew University, Jerusalem, Israel 91904

## Abstract

Magnetization measurements were carried out in the in field-cooled (FC) and in zero-field-cooled (ZFC) conditions versus temperature, T, and external magnetic field, H, on a sigma-phase $Fe_{47}Mo_{53}$ compound. Analysis of the measured $M_{FC}$ and $M_{ZFC}$ curves yielded values of characteristic temperatures: magnetic ordering (Curie) temperature, $T_C$, irreversibility temperature, $T_{ir}$, temperature of the maximum in $M_{ZFC}$, $T_m$, identified as the Nèel ($T_N$) temperature, and cross-over temperature, $T_{co}$. Based on these temperatures a magnetic phase diagram in the H-T plane was outlined. The field dependences of the characteristic temperatures viz. of the irreversibility and of the cross-over temperatures were described in terms of a power law with the exponent 0.5(1). In the whole range of H i.e. up to 800 Oe, except the one H $\leq$ 50 Oe, a rare double re-entrant transition viz. PM→FM→AF→SG takes place. For small fields i.e. H $\leq$ 50 Oe a single re-entrant transition viz. PM →FM→SG is revealed.

*Corresponding author: Stanislaw.Dubiel@fis.agh.edu.pl

# 1. Introduction

The sigma phase ($\sigma$) belongs to a family of the so-called Frank-Kasper phases, also known as topologically closed-packed (TCP) structures. It has a complex tetragonal crystallographic structure. Its unit cell, whose volume is of the order of ~350 Å$^3$, hosts 30 atoms distributed over 5 nonequivalent lattice sites having high (12-15) coordination numbers. The $\sigma$ phase occurs in alloys in which at least one element is a transition metal. In the simplest alloys i.e. binary ones there are known 43 cases in which $\sigma$ was revealed [1]. Concerning magnetic properties of $\sigma$, a subject of the present study, they were for the first time detected for $\sigma$-FeV [2] and a few years later for $\sigma$-FeCr systems [3]. For about five decades the magnetic ordering in these two cases was regarded as ferromagnetic (FM). However, it has turned out that the magnetism in these two alloy systems is much more complex. Namely, a re-entrant character was revealed i.e. the FM state turned out be an intermediate one while a spin-glass (SG) state was found to be the ground state [4]. In recent years magnetism of $\sigma$ was discovered in other two binary alloys of Fe viz. Fe-Re [5] and Fe-Mo [6,7]. Alike in the Fe-V and Fe-Cr systems, the $\sigma$-FeRe and $\sigma$-FeMo compounds can be assigned as low-temperature, weak and highly itinerant FM, which exhibit re-entrant properties. However, contrary to $\sigma$-FeV and $\sigma$-FeCr cases for which the field-cooled (FC) magnetization curves showed the Brillouin-like behavior [4], for both $\sigma$-FeRe [5] and $\sigma$-FeMo [7] the $M_{FC}$ curves have a maximum followed by a concave-like shape at lower temperatures. The latter, is in line with an antiferromagnetic (AF) behavior. To shed more light and get a deeper insight into the magnetism of $\sigma$ in the Fe-Mo system, a systematic DC magnetization study versus temperature, T, and applied magnetic field, H, was performed on a $\sigma$-Fe$_{47}$Mo$_{53}$ alloy. Based on the obtained results, a magnetic phase diagram in the H-T plane has been outlined and presented in this paper.

# 2. Experimental

## 2.1. Fe$_{47}$Mo$_{53}$ sample preparation

The present study was carried out on one of sigma-phase samples of Fe$_{100-x}$Mo$_x$, viz. x=53, that were used in previous investigations [6,7,8]. It was produced by sintering of iron (99.9% purity) and molybdenum (99.95% purity) powders. The constituents were first mixed in the adequate proportion and next compressed into pellets which were vacuum annealed at 1700 K for 6 h followed by quenching into liquid nitrogen. Powder X-ray diffraction clearly indicates a tetragonal structure, which is one characteristic of the sigma phase [8].

## 2.2. Magnetization measurements

Magnetization (M) measurements were carried out using a commercial (Quantum Design) superconducting quantum interference device (SQUID) magnetometer. The sample was mounted in gel-cap. Prior to recording the zero-field-cooled magnetization curves, $M_{ZFC}$, the SQUID magnetometer was always adjusted to be in a *"true"* H = 0 state. The temperature dependence of the field-cooled ($M_{FC}$) and the ZFC branches were taken via warming from 5 K to 45 K. Both $M_{FC}$ and $M_{ZFC}$ branches were recorded in applied magnetic field, ranging between 25 and 800 Oe. The measurements are displayed in Fig. 1.

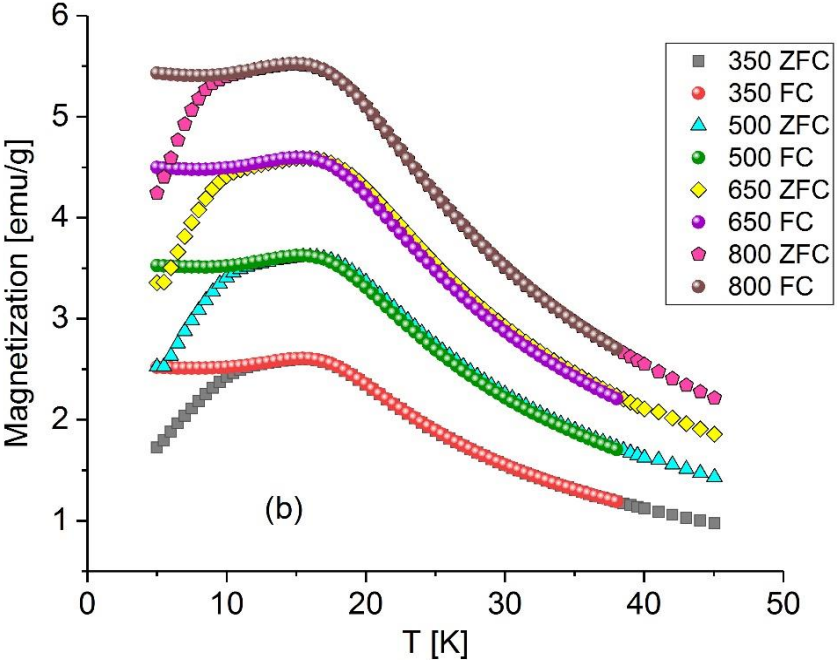

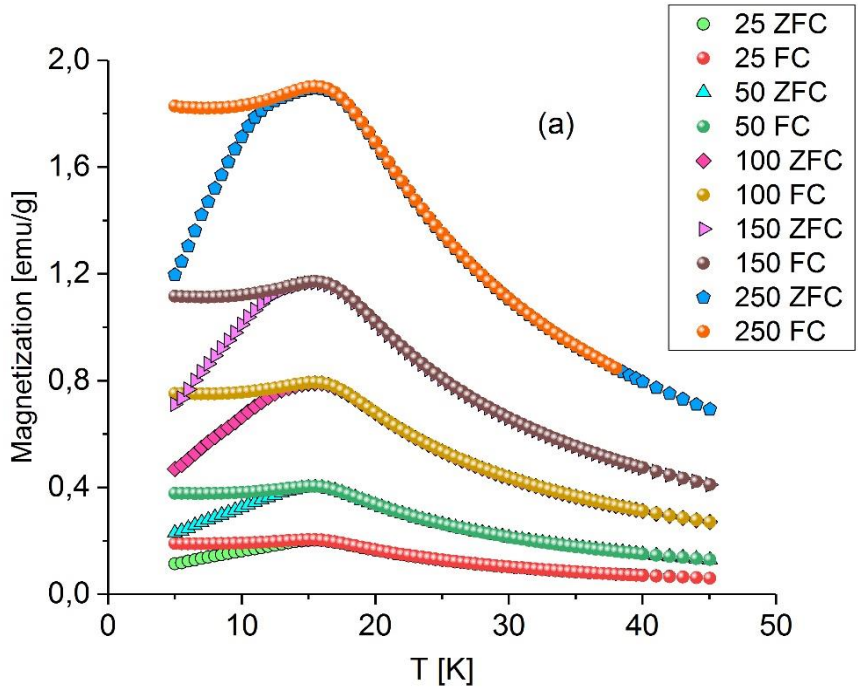

Fig. 1 Magnetization curves vs. temperature, ~~T,~~ measured in the field-cooled (FC) and in zero-field-cooled (ZFC) conditions. Values of the applied field are given in Oe.

## 3. Results and discussion

### 3.1. Characteristic temperatures

Three to four different characteristic temperatures can be distinguished in order to describe the temperature dependence of the $M_{ZFC}$ and $M_{FC}$ curves. Namely, going from high to low T-range, we can identify: (1) the magnetic ordering temperatures, to a FM state, $T_C$, (2) the irreversibility (bifurcation) temperatures, $T_{ir}$, (3) the temperature of maximum values in $M_{ZFC}$, $T_m$, and (4) the knee temperatures in $M_{ZFC}$, $T_{co}$. Approximate positions of these temperatures are indicated by arrows in Figs. 2-4. The interpretation for re-entrant spin-glasses is as follows: $T_C$ marks the Curie temperature transition from a paramagnetic (PM) state into a FM one, (or $T_N$ for an antiferromagnetic one,), $T_{ir}$ indicates a transition from the FM (AF) state into a spin-glass (SG) state with a weak irreversibility, $T_m$ (if lower than $T_{ir}$) indicates a transition from this SG-state into a strong irreversibility SG state, which is usually defined as a cross-over temperature, $T_{co}$. The latter case can be seen in the presently studied sample for H < 50 Oe, as illustrated in Fig. 2a. For H ≥100 Oe $T_{ir} < T_m$, see Fig. 2b. In this case the $M_{ZFC}$ curves have a "knee" at $T_{co}$. $T_C$ can be precisely determined as the position of the inflection points of the derivative of the $M_{FC}$ or $M_{ZFC}$ curves as illustrated in Fig. 4. All these temperatures when measured for different H-values can be used to construct a magnetic phase diagram in the H-T coordinates.

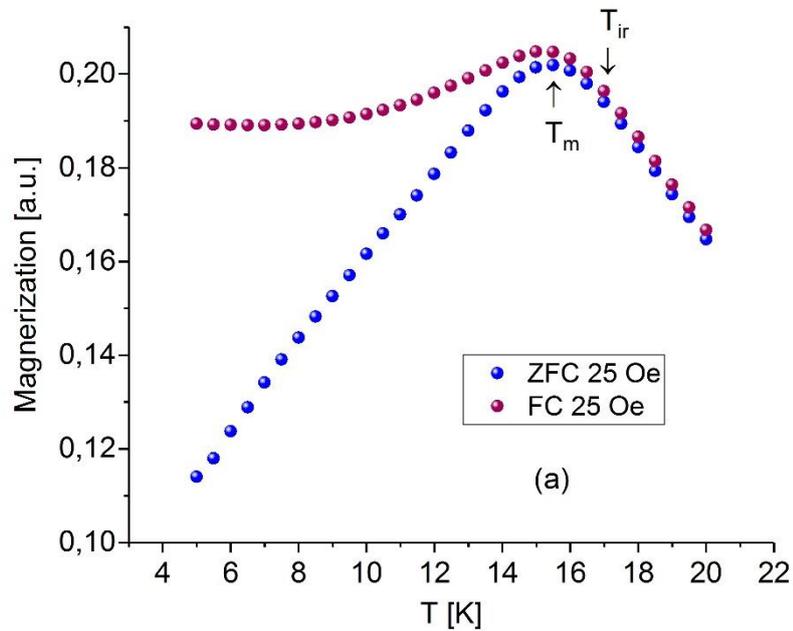

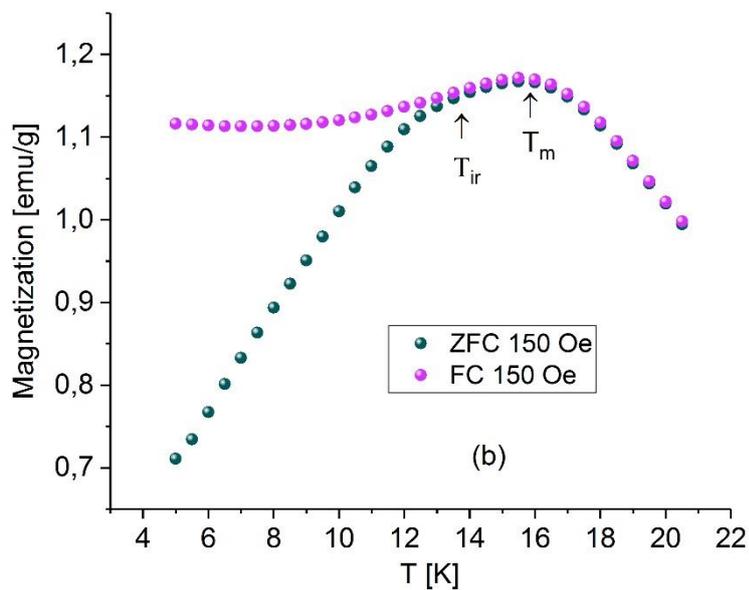

Fig.2 FC and ZFC magnetization curves in a low temperature range measured in (a) H=25 Oe and (b) H=150 Oe. Arrows indicate positions of the maximum, $T_m$, and the irreversibility, $T_{ir}$ temperatures. Note that for (a) $T_{ir} > T_m$ and for (b) $T_{ir} < T_m$.

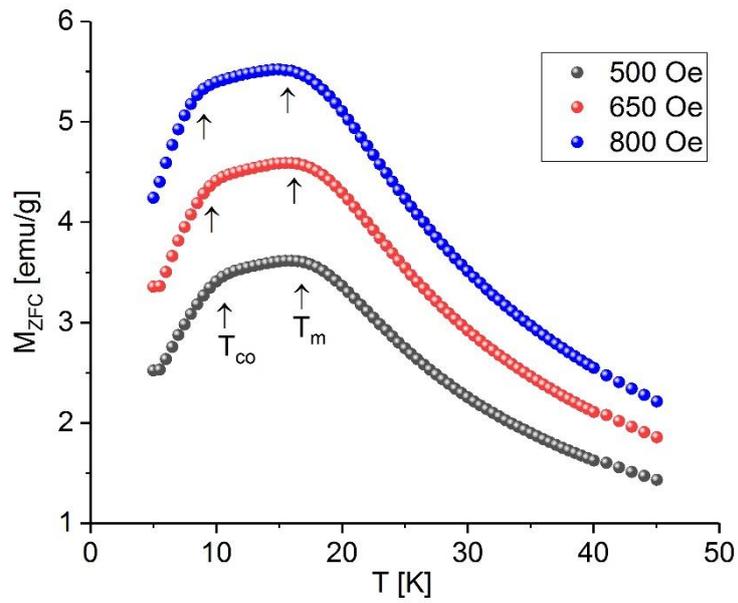

Fig. 3 Examples of the ZFC curves measured in various fields shown in the legend. Arrows mark positions of the maxima, $T_m$, and the knees, $T_{co}$ temperatures.

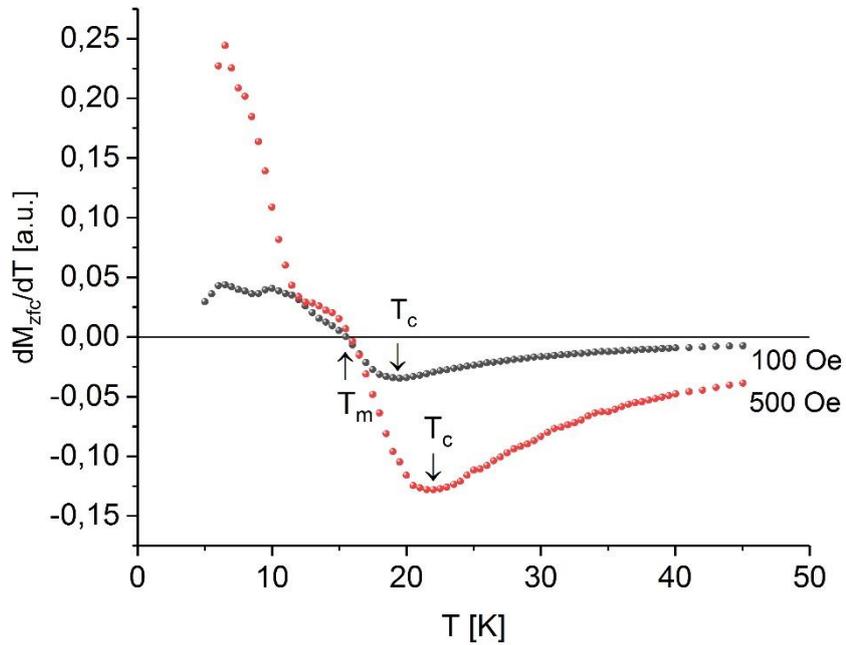

Fig. 4  Temperature derivatives of the $M_{ZFC}$ curves recorded for 100 and 250 Oe . . Positions of the Curie temperature, $T_C$, and that of the maxima, $T_m$, are indicated by arrows.

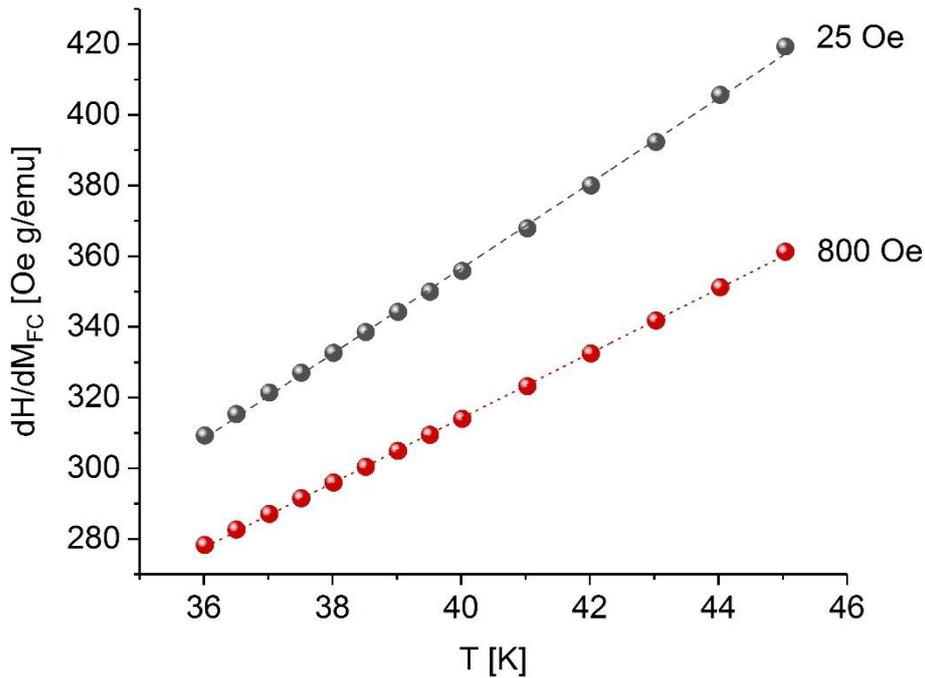

Fig. 5 Linear part of the reciprocal of the magnetic susceptibility vs. temperature for $Fe_{47}Mo_{53}$ measured above Tc under applied fields of 25 and 800 Oe. The straight lines represent the best fits to data in terms of the generalized Curie-Weiss law.

$T_C$ obtained is consistent with the positive values of the Curie-Weiss temperature, $\Theta_{CW}$. Indeed, as presented in Fig. 5 for H=25 Oe and 800 Oe, the reciprocal DC susceptibility, $dH/dM_{ZFC}$, could be well-fitted in the paramagnetic state (PM) (above Tc) to the generalized Curie-Weiss law yielding $\Theta_{CW}$=10.4 K for the lowest field and 5.1 K for the highest one. All obtained $\Theta_{CW}$-values are plotted vs. H are shown in Fig. 6 together with the derived therefrom frustration degree values given by, FD=$T_C/\Theta_{CW}$ (Values of $T_C$ are shown in Fig. 11).

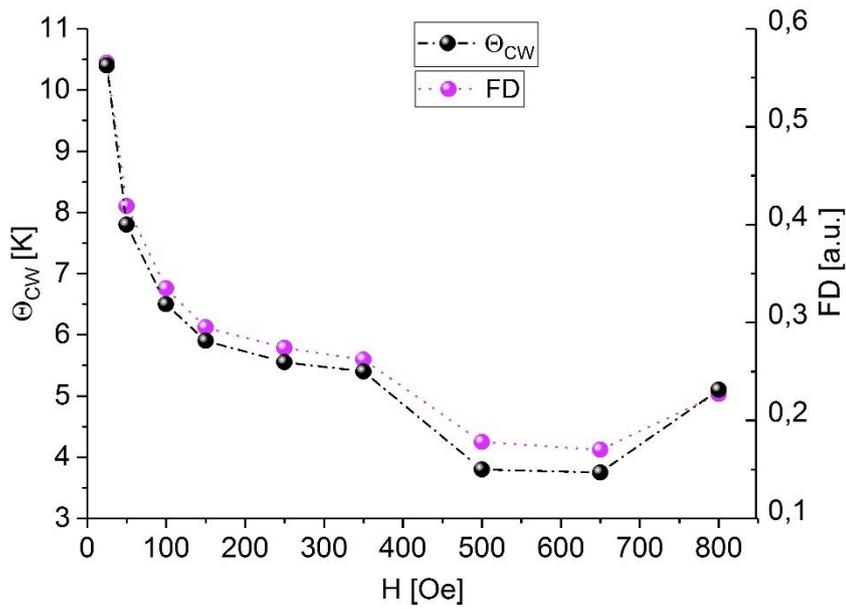

Fig. 6 Field dependence of the Curie-Weiss temperature, $\Theta_{CW}$, and of the frustration degree, FD.

One can also easily and precisely determine the position of the maxima in $M_{ZFC}$, $T_m$, as at $T=T_m$ $dM_{ZFC}/dT=0$ – see Fig. 4. Determination of $T_{ir}$ is more complicated. Usually one may plot the differences $\Delta M=M_{FC}-M_{ZFC}$ as a function of temperature, as shown in Fig. 7, and determines the value of $T_{ir}$ as the one below which $\Delta M \geq 0$ e. g. [9].

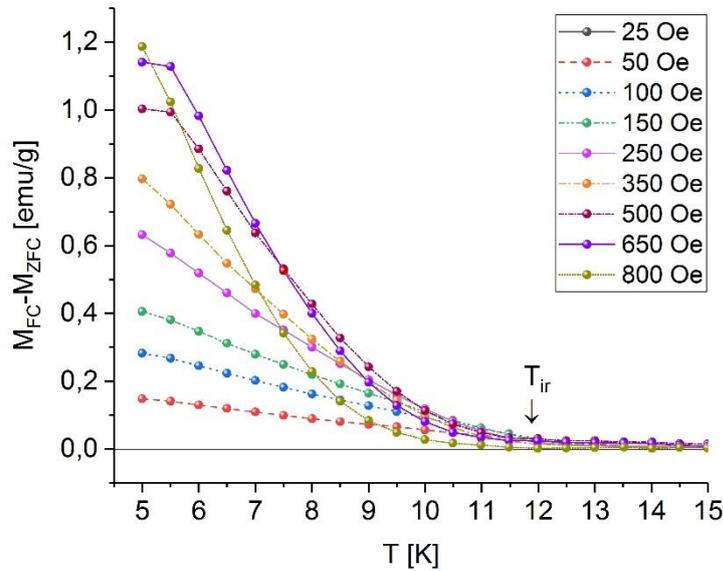

Fig. 7 Difference in the FC and ZFC magnetization curves vs. temperature as obtained for all values of the magnetic field shown in the legend. Approximate position of the irreversibility temperature, $T_{ir}$, is marked by arrow.

The same plots shown in Fig. 7, can be sometimes, but not always, used for determination of $T_{co}$. Namely, in some cases the $\Delta M(T)$ curves show, at a pretty-well defined temperature, a steeper increase. These temperatures, often determined by a tangent or double-tangent method, are regarded as $T_{co}$ e. g. [9,10]. However, sometimes the $\Delta M(T)$ curves do not exhibit any well-defined anomaly. Alternatively, $T_{co}$ can be determined from the "knee" in the $M_{ZFC}$ curve applying a double-tangent method – see e. g. [11]. However, in our opinion this approach cannot give a unique estimation of $T_{co}$ because the $M_{ZFC}(T)$ dependence is neither linear with T above nor below the "knee". Instead, in the present analysis of the $M_{ZFC}$ curves we based our determination of $T_{co}$ on the observation that the second temperature derivative, $d^2M_{ZFC}/dT^2$ had a minimum at $T < T_{ir}$ for all values of H – see an example in Fig. 8. Thus, in the present case, we have associated $T_{co}$ with this minimum.

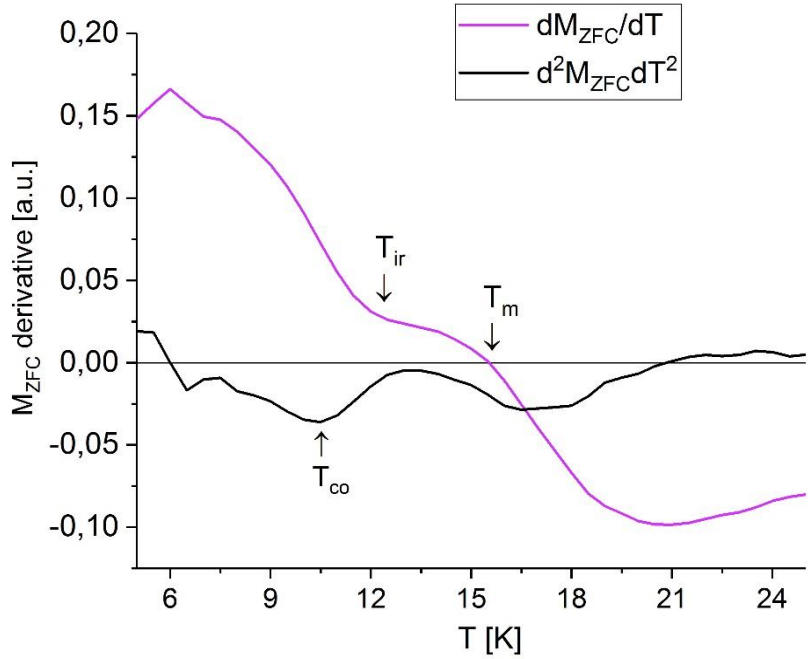

Fig. 8 First and second temperature derivatives of $M_{ZFC}$ recorded in the magnetic field of 350 Oe. Positions of 3 characteristic temperatures are marked by arrows.

## 3.2. Irreversibility

It is evident from Fig. 1 that the difference between the $M_{FC}$ and $M_{ZFC}$ curves, $\Delta M$, depends on the applied field. As $\Delta M$ can be regarded as a measure of irreversibility, it is of interest to see how it changes with H. In the case of σ-phase Fe-V and Fe-Cr alloys the $M_{FC}$ curves resembled the Brillouin-like function and $\Delta M$, was decreasing with H, and tending to zero [4]. In the present case, the $M_{FC}$ curves are different, namely they have a maximum at the same temperature as the $M_{ZFC}$ curves i.e. $T_m$, and at lower T their shape is concave.

In order to quantitatively describe the irreversibility we calculated the difference in area under the $M_{FC}(T,H)$ and $M_{ZFC}(T,H)$ curves, $\Delta M$ (H):

$$\Delta M(H) = \int_{T_1}^{T_2} (M_{FC}(H) - M_{ZFC}(H))dT \qquad (1)$$

Where $T_1$=5 K and $T_2=T_{ir}$. Figure 9 shows the obtained result;

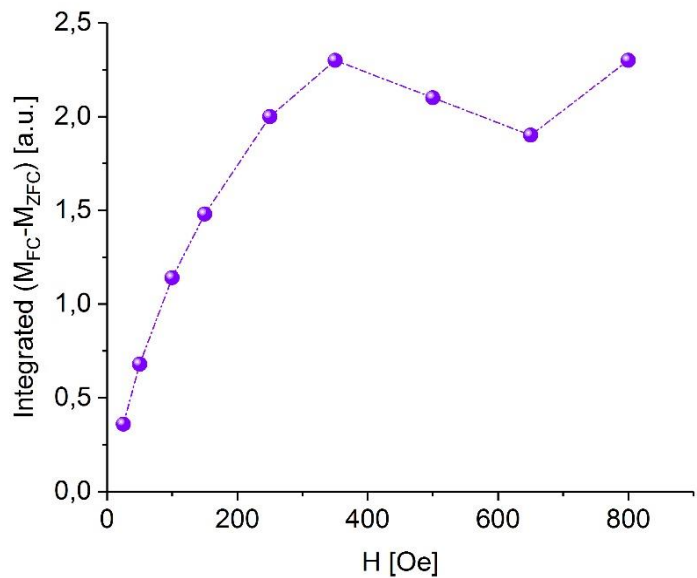

Fig. 9  Temperature integrated difference between the FC and ZFC magnetization curves vs. the applied magnetic field, H. The dashed line is to guide the eye.

As follows from Fig. 9, $\Delta M$ is not a monotonous function of H since it has a maximum at H=350 Oe and a minimum at H=650 Oe. This reflects a complex magnetic behavior of the sample.

## 3.4. Magnetic phase diagram

Figure 10 clearly demonstrates that the applied magnetic field significantly affects the magnetization curves and the values of the characteristic temperatures. In particular we can easily notice that while for H=25 Oe $T_{ir} > T_m$ and the $M_{ZFC}$ curve does not have a visible "knee", for H=800 Oe $T_{ir} < T_m$ and the $M_{ZFC}$ curve displays a pronounced "knee". These features give qualitative evidence that the magnetism of re-entrant SGs significantly depends on H.

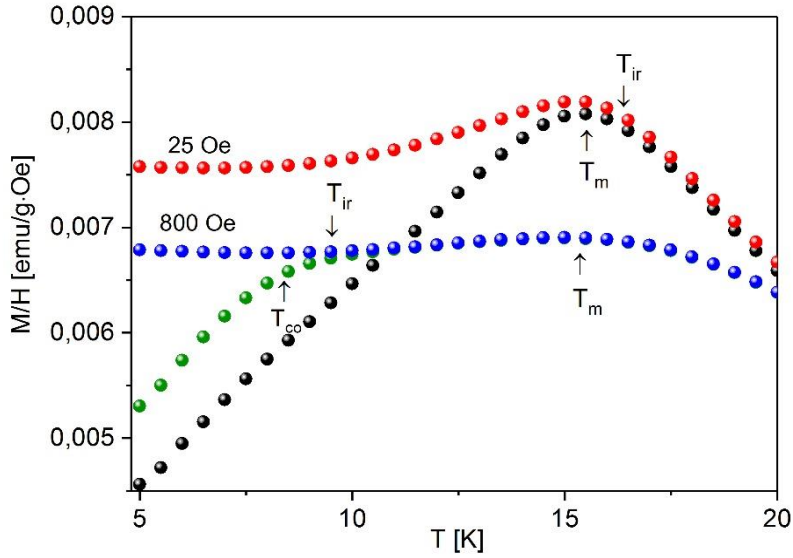

Fig. 10 DC magnetic susceptibility, M/H, vs. temperature, illustrating an effect of the applied magnetic field (25 and 800 Oe) on positions of the irreversibility and the knee.

A quantitative picture showing possible magnetic states can be obtained by constructing a magnetic phase diagram in the H-T coordinates. To this end we have presented in Fig. 11 H-dependencies of the characteristic temperatures ~~we have~~ determined as described in Paragraph 3.1. First of all we have to notice that the magnetism of $Fe_{47}Mo_{53}$ is really complex as for H > 50 Oe we can see a double re-entrant transition viz. PM→FM→AF→SG which is very rare. Whereas, for low fields i.e. H ≤ 50 Oe a single re-entrant transition exists viz. PM→FM→SG. For all H, however, the SG state can be divided into two sub states: SG1 with a low irreversibility and SG2 with the strong one. For re-entrant SGs the mean-field theory, and, in particular, the Gabay-Toulouse (GT) model [12] predict the existence of two characteristic lines which are regarded as the border lines between the two sub states with different degree of irreversibility. The one between the FM phase and the SG1 state can be experimentally constructed as *loci* of $T_{ir}(H)$, whereas the border line between SG1 and SG2 as *loci* of $T_{co}(H)$. Following the GT-model, the H-dependence of both these line can be described by a power law:

$$T(H) = T(0) - a \cdot H^{\varphi} \qquad (1)$$

Where T can be either $T_{ir}$ or $T_{co}$.

Predicted values of $\varphi$ are: 2/3 for $T_{ir}$ and 2 for $T_{co}$ [12]. Within the uncertainties, in the present case for both temperatures $\varphi \approx 0.5$, hence the agreement with the GT model is merely qualitative.

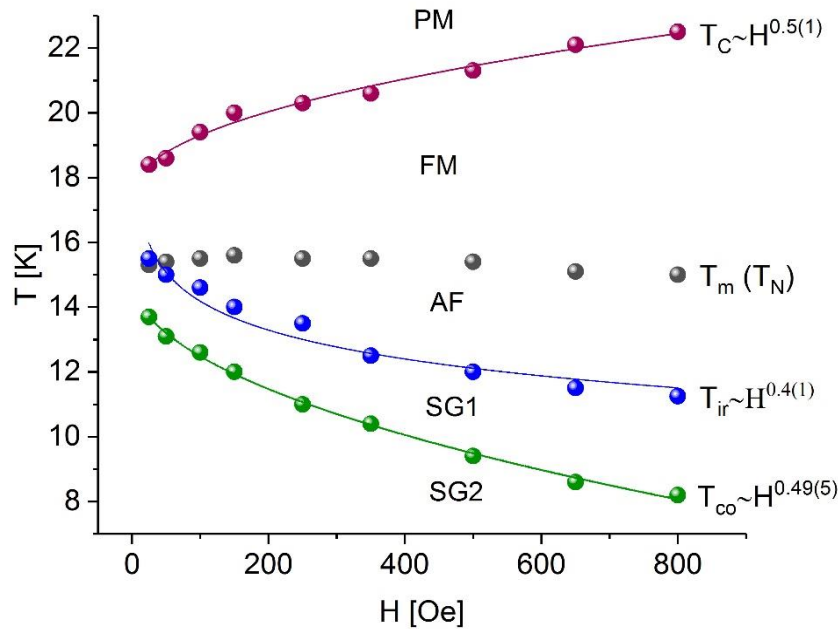

Fig. 11 The magnetic phase diagram in the H-T coordinates as determined for the sigma-phase $Fe_{47}Mo_{53}$ compound based on the present study. The meaning of the symbols and of acronyms is given in the text.

The complex magnetic structure of $Fe_{47}Mo_{53}$ stems from the complex crystallographic structure of $\sigma$ i.e. five different lattice sites that in the case of binary alloys both elements distribute randomly over the five sites.. According to theoretical calculations for magnetic $\sigma$-phase systems, both elements have different magnetic moments [5,13-16], and in some cases their orientation may be parallel or anti-parallel, depending on the crystallographic site and alloy composition. The latter case was reported for Fe based $\sigma$ in Fe-Cr [13], Fe-V [15], Fe-Re [5] and Fe-Mo [16] alloys systems. It should be remembered that calculations mentioned above were assumed only linear ordering of the moments at T= 0 K. At elevated temperatures under external magnetic field the energy relations of the systems are likely different, hence the moments magnitudes and orientations are also different. Indeed, the only calculations, we are aware of, that were performed for elevated temperatures predict for σ-FeCr, that both Fe and Cr moments are temperature dependent [14]. Unfortunately, experimental determination of magnetic moments associated with particular lattice sites is practically hardly possible (at least no such reports are available in the literature hence validation of the calculations cannot be done. The only paper relevant to the site-resolved moments is the one in which $^{61}$V nuclear magnetic resonance spectra were recorded for σ-FeV alloys based on which magnetic moments of V atoms were estimated [17].

## 4. Conclusions

On the grounds of the present study, a magnetic phase diagram of a sigma-phase Fe$_{47}$Mo$_{53}$ alloy has been proposed. It gives a quantitative picture of the effect of the applied magnetic field on the magnetic states of the investigated samples. Going from high to low temperatures, the obtained results can be concluded as follows:

1. For H ≤ 50 Oe there is a normal re-entrant transition from a paramagnetic (PM) to a spin-glass (SG) state over an intermediate ferromagnetic (FM) state.

2. For H > 50 Oe there is a double re-entrant transition PM→FM→AF→SG. In other words there are two intermediate states between the PM and the ground state.

3. The SG state is heterogeneous i.e. it can be divided into two sub states: SG1 with a weak irreversibility and SG2 with a strong irreversibility.

4. The H-dependences of the irreversibility and of the cross-over temperatures agree qualitatively with the Gabay-Toulouse model.

## Acknowledgement

This work was financed by the Faculty of Physics and Applied Computer Science AGH UST statutory tasks within subsidy of Ministry of Science and Higher Education in Warsaw.